\documentclass[11pt,epsf,letterpaper]{article}
\usepackage{setspace}
\usepackage{color}
\usepackage{amsmath}
\usepackage{amsfonts}
\usepackage{verbatim}
\usepackage{amssymb}
\usepackage{graphicx,bm}
\usepackage{graphicx}
\usepackage{graphicx}
\usepackage{subcaption}
\usepackage{epstopdf}

\definecolor{darkgreen}{rgb}{0,0.35,0}

\providecommand{\U}[1]{\protect\rule{.1in}{.1in}}
\begin{document}

\title{Integrability and chemical potential \\
in the (3+1)-dimensional Skyrme model}
\author{P. D. Alvarez$^{1}$, F. Canfora$^{2}$, N. Dimakis$^{3}$, A.
Paliathanasis$^{3,4}$ \\
$^{1}$\textit{Universidad T\'ecnica Federico Santa Mar\'ia, Santiago, Chile.}%
\\
$^{2}$\textit{Centro de Estudios Cient\'{\i}ficos (CECS), Casilla 1469,
Valdivia, Chile.}\\
$^{3}$\textit{Instituto de Ciencias Fisicas y Matematicas, Universidad
Austral de Chile, Valdivia, Chile.}\\
$^{4}$\textit{Institute of Systems Science, Durban University of Technology,
PO Box 1334, }\\
\textit{Durban 4000, Republic of South Africa.}\\
{\small pedro.alvarezn@usm.cl, canfora@cecs.cl, nsdimakis@gmail.com ,
anpaliat@phys.uoa.gr}}
\maketitle

\begin{abstract}
Using a remarkable mapping from the original (3+1)dimensional Skyrme model
to the Sine-Gordon model, we construct the first analytic examples of
Skyrmions as well as of Skyrmions--anti-Skyrmions bound states within a
finite box in 3+1 dimensional flat space-time. An analytic upper bound on
the number of these Skyrmions--anti-Skyrmions bound states is derived. We
compute the critical isospin chemical potential beyond which these Skyrmions
cease to exist. With these tools, we also construct topologically protected
time-crystals: time-periodic configurations whose time-dependence is
protected by their non-trivial winding number. These are striking
realizations of the ideas of Shapere and Wilczek. The critical isospin
chemical potential for these time-crystals is determined.
\end{abstract}

\section{Introduction}

One of the most beautiful examples of the interplay between topology and
field theory is provided by the Skyrme theory \cite{skyrme}. Its importance
in nuclear and particle physics (for instance see \cite%
{nuc0,nuc1,nuc2,nuc3,nuc4,nuc5}) arises from its relations with low energy
QCD \cite{witten0} and it is the prototype of a non-integrable model. The
non-trivial role of topology appears since the Skyrme model supports
solitons, called \textit{Skyrmions}, which are topologically stable and
represent Baryonic degrees of freedom (see \cite{finkrub}\ \cite{manton}
\cite{skyrev1} \cite{witten0} \cite{giulini} \cite{bala0} \cite{ANW} \cite%
{guada} and references therein). The identification of the Baryon number in
particle physics with the third homotopy class of the Skyrmion \cite{witten0}
showed that the original Skyrme intuition was correct.

Morover, the ideas of Skyrme have nontrivial applications not only in
particles and nuclear physics but also in many other areas of physics. In
\cite{sklat} it was found that skyrmion lattices exist in semiconductors.
What is more, in an antiferromagnetic spinor Bose-Einstein condensate a
two-dimensional skyrmion was observed \cite{sklat2} to be stable on a short
time. Applications of Skyrme model are also encountered in gravitational
physics, for instance in cosmology \cite{con01,con02} or black holes physics
where it has been found that the \textquotedblleft no
hair\textquotedblright\ conjecture can be violated \cite%
{bh01,bh03,bh03b,Ioannidou,bh04,bh05}. A further field of research in which
the role of Skyrmions is extremely relevant is the analysis of magnetic
materials (a nice review being \cite{MagSkyrme}).

The original Skyrme model is very far from being integrable and only very
few explicit analytic results are known. In particular, there is no explicit
solution with non-vanishing Baryon number on flat space-times. Consequently,
the Skyrme phase diagrams (which could provide very valuable informations on
nuclear matter) are very difficult to approach with analytical methods.

By using the original spherical hedgehog ansatz of Skyrme, Klebanov proposed
a phenomenological approach to the analysis of Skyrmions at finite density a
long time ago \cite{klebanov}. By following his point of view and using the
technique of \cite{chemical1} \cite{chemical2}, a non-vanishing isospin
chemical potential was introduced in \cite{chemical3} \cite{chemical4}.
However, both finite volume effects and isospin chemical potential break
spherical symmetry, and this fact makes difficult to apply the spherical
hedgehog ansatz in these cases. Without an appropriate ansatz (which is both
topologically non-trivial and non-spherically symmetric) it becomes very
hard to derive analytic results on how the Isospin chemical potential as
well as finite-volume effects affect the behavior of Skyrmions.

In recent years, a generalized hedgehog ansatz was proposed, which provides
with a strategy to construct a non-spherical hedgehog-like ansatz with the
right properties (see \cite{canfora1,canfora2,canfora3,canfora4}, \cite%
{yang1}, \cite{canfora6}, \cite{cantalla3}, \cite{cantalla3a}, \cite%
{cantalla3b}, \cite{cantalla4} and references therein) both in the Skyrme
and in the Yang-Mills-Higgs cases. By using this technique, we construct
analytic and topologically non-trivial solutions of the Skyrme model without
spherical symmetry living within a finite box in flat space-times. We also
construct analytic Skyrmions--anti-Skyrmions bound states. We derive a bound
on the number of Skyrmions--anti-Skyrmions bound states in terms of the
coupling constant. The isospin chemical potential can be included keeping,
at the same time, the nice properties of the ansatz. The critical isospin
chemical potential, beyond which the Skyrmion living in the box ceases to
exist, can be explicitly determined.

Remarkably, the generalized hedgehog ansatz allowed us to construct novel
types of topological configurations of the Skyrme model that can be defined
as \textit{topologically protected time crystals}, see below for more
information.

The idea of time crystal was introduced by Wilczek and Shapere in \cite%
{timec1} \cite{timec2} \cite{timec3}, based on the following observations.
\textit{Spontaneous Symmetry Breaking} is a general property of nature that
manifests itself in many different situations. It refers to situations where
the observed configurations of a given system possess less symmetries than
the corresponding action. They proposed the following very interesting and
intriguing question: \textit{is it possible to break spontaneously time
translation symmetry?}

Powerful no-go theorems \cite{timec4} \cite{timec5} ruled out the original
proposals but they inspired a huge number of physicists to open new research
lines (a nice review is \cite{timecr}). New types of time crystals in
condensed matter physics have been proposed and realized in laboratories
\cite{timec5.5} \cite{timec5.6} \cite{timec5.7} \cite{timec6} \cite{timec7}
\cite{timec9} (an up-to-date list of references can be found in \cite{timecr}%
). However, no example in nuclear and particles physics has been considered
so far.

Here we show explicitly that, when finite volume effects are taken into
account, the Skyrme model predicts the existence of a new type of
time-crystal. These are exact time-periodic configurations of the
(3+1)-dimensional Skyrme model that cannot be deformed continuously to the
trivial vacuum as they possess a non-trivial winding number extending along
the time-direction (which is a sort of Lorentzian version of the Euclidean
instanton number). Due to topological reasons, these time crystals can only
decay into other time-periodic configurations: correspondingly, the
time-periodicity is topologically protected. Hence the name \textit{%
topologically protected time crystals}.

This paper is organized as follows: in section \ref{model} we introduce the
Skyrme action. In section \ref{sine-gordon}, we discuss the sine-Gordon
mapping and the effects of the chemical potential. In section \ref{time
crystal}, we describe the topologically protected time-crystals. In section %
\ref{conclusions}, we draw some concluding ideas.

\section{The Skyrme Model}

\label{model}

We consider the $SU(2)$ Skyrme model in four dimensions. The action of the
system is
\begin{align}
S& =\frac{K}{2}\int d^{4}x\sqrt{-g}\left[ \frac{1}{2}\mathrm{Tr}\left(
R^{\mu }R_{\mu }\right) +\frac{\lambda }{16}\mathrm{Tr}\left( G_{\mu \nu
}G^{\mu \nu }\right) \right] \ ,  \label{sky1} \\
R_{\mu }& =U^{-1}\nabla _{\mu }U\ ,\ \ G_{\mu \nu }=\left[ R_{\mu },R_{\nu }%
\right] \ ,  \label{sky2} \\
U& \in SU(2)\ ,\ \ R_{\mu }=R_{\mu }^{j}t_{j}\ ,\ \ t_{j}=i\sigma _{j}\ ,
\label{sky2.5}
\end{align}%
where $\sqrt{-g}$ is the (square root of minus) the determinant of the
metric, the positive parameters $K$ and $\lambda $ are fixed experimentally
and $\sigma _{j}$ are the Pauli matrices. In our conventions $c=\hbar =1$,
the space-time signature is $(-,+,+,+)$ and Greek indices run over
space-time. The stress-energy tensor is
\begin{equation}
T_{\mu \nu } =-\frac{K}{2}\mathrm{Tr}\left[ R_{\mu }R_{\nu }-\frac{1}{2}%
g_{\mu \nu }R^{\alpha }R_{\alpha }\right. \, \left. \frac{\lambda }{4}\left(
g^{\alpha \beta }G_{\mu \alpha }G_{\nu \beta }-\frac{g_{\mu \nu }}{4}%
G_{\sigma \rho }G^{\sigma \rho }\right) \right] \ ,  \notag  \label{timunu1}
\end{equation}
and the matter field equations are
\begin{equation}
\nabla ^{\mu }\left( R_{\mu }+\frac{\lambda }{4}\left[ R^{\nu },G_{\mu \nu }%
\right] \right) =0.  \label{nonlinearsigma1}
\end{equation}

We adopt a standard parametrization of the $SU(2)$-valued scalar $U(x^{\mu
}) $
\begin{equation}
U^{\pm 1}(x^{\mu })=Y^{0}(x^{\mu })\mathbb{\mathbf{I}}\pm Y^{i}(x^{\mu
})t_{i}\ ,\ \ \left( Y^{0}\right) ^{2}+Y^{i}Y_{i}=1\,,  \label{standnorm}
\end{equation}%
where $\mathbb{\mathbf{I}}$ is the $2\times 2$ identity and
\begin{align}
Y^{0}& =\cos C\ ,\ Y^{i}=n^{i}\cdot \sin C\ ,  \label{pions1} \\
n^{1}& =\sin F\cos G\ ,\ \ n^{2}=\sin F\sin G\ ,n^{3}=\cos F\ .
\label{pions2}
\end{align}

The Skyrme field possesses a non-trivial topological charge which,
mathematically, is a suitable homotopy class or winding number: its explicit
expression as an integral over a suitable three-dimensional hypersurface $%
\Sigma $ is
\begin{equation}
W=-\frac{1}{24\pi ^{2}}\int_{\Sigma }\epsilon ^{ijk}Tr\left( U^{-1}\partial
_{i}U\right) \left( U^{-1}\partial _{j}U\right) \left( U^{-1}\partial
_{k}U\right) =-\frac{1}{24\pi ^{2}}\int_{\Sigma }\rho _{B}\ ,  \label{new4}
\end{equation}%
where the baryon density is defined by $\rho _{B}=12\sin ^{2}C\sin F\
dC\wedge dF\wedge dG$. A necessary condition to have a non-vanishing baryon
density is $dC\wedge dF\wedge dG\neq 0$.

When, in the above integral, the three-dimensional hypersurface $\Sigma $ is
space-like then the topological charge is interpreted as Baryon number.
However, due to the fact that $\rho _{B}$ does not depend on the metric,
there are two further options: $\Sigma $ can be time-like or light-like. The
last two possibilities have not been explored so far in the literature. In
fact they are extremely interesting as whenever $W\neq 0$ (whether $\Sigma $
is space-like, time-like or light-like) the corresponding Skyrme
configuration has a non-trivial homotopy and, consequently, cannot be
deformed continuously into the trivial vacuum $U=\mathbb{\mathbf{I}}$. The
cases in which $\Sigma $ is time-like and $W\neq 0$ correspond to
topologically protected time crystals as it will be explained below. We will
only consider an ansatz in which $\rho _{B}\neq 0$.

The natural generalization of the hedgehog ansatz introduced in \cite%
{canfora6} in the cases in which the metric is flat reads
\begin{equation}  \label{pions2.25}
G =\frac{\gamma +\phi }{2\,}\ ,\ \ \tan F=\frac{\tan H}{\cos A}\ ,\ \ \tan C=%
\frac{\sqrt{1+\tan ^{2}F}}{\tan A}\ ,
\end{equation}
where
\begin{equation}  \label{pions2.26}
A =\frac{\gamma -\phi }{2\,}\ ,\ \ H=H\left( r,z\right) \ .
\end{equation}

It can be verified directly that, the topological density $\rho _{B}$ is
non-vanishing. From the standard parametrization of $SU(2)$ (\cite{Shnir})
it follows that
\begin{equation}
0\leq \gamma \leq 4\pi ,\quad 0\leq \phi \leq 2\pi \ ,  \label{domain}
\end{equation}%
while the boundary condition for $H$ will be discussed below.

\section{Sine-Gordon and Skyrmions}

\label{sine-gordon}

Let us consider the following flat metric
\begin{equation}
ds^{2}=-dz^{2}+\ell ^{2}\left( dr^{2}+d\gamma ^{2}+d\phi ^{2}\right) \ ,
\label{Minkowski}
\end{equation}%
(in this section $z$ is the time variable). The length $\ell $\ represents
the size of the box where the Skyrmion lives. The coordinates $r$, $\gamma $
and $\phi $ are angular coordinates; the domain of $\gamma$ and $\phi$ is
given by \eqref{domain}, while for $r$ we choose the finite interval $0\leq
r \leq 2\pi$.

The full Skyrme field equations (\ref{nonlinearsigma1}) with the generalized
hedgehog ansatz in Eqs. (\ref{pions1}), (\ref{pions2}), (\ref{pions2.25})
and (\ref{pions2.26}) \textit{reduce to just one scalar differential
equation for the profile} $H$
\begin{equation}
\Box H-\frac{\lambda }{8\,\ell ^{2}(\lambda +2\ell ^{2})}\sin \left(
4H\right) =0\ ,  \label{sineG}
\end{equation}%
where $\Box $\ is the two-dimensional D' Alambert operator.

The energy of the configuration is given by
\begin{equation}
E=\int \!\! \ell ^{3} T^{00}drd\gamma d\phi \,,  \label{energy}
\end{equation}%
where
\begin{equation}
T_{00}=\frac{K}{64\,\ell ^{4}}\left[ 16(\lambda +2\,\ell ^{2})\left(
(\partial _{r}H)^{2}+\ell ^{2}(\partial _{z}H)^{2}\right) +\lambda \left(
1-\cos (4H)\right) +16\,\ell ^{2}\right] \,.  \label{T00a}
\end{equation}

The topological Baryon charge $B$ and charge density $\rho _{B}$ become
respectively
\begin{equation}
B=-\frac{1}{24\pi ^{2}}\int_{t=const}\rho _{B}\,,\ \rho _{B}=3\sin
(2H)dHd\gamma d\phi \ .  \label{td1}
\end{equation}

If we replace the topologically non-trivial ansatz in Eqs. (\ref{pions1}), (%
\ref{pions2}), (\ref{pions2.25}) and (\ref{pions2.26}) into the original
action (\ref{sky1}) we obtain an effective action given by
\begin{equation}
\mathcal{L}(H)=16\,\ell ^{2}(\lambda +2\ell ^{2})\nabla _{\mu }H\nabla ^{\mu
}H-\lambda \cos (4H),  \label{LagSG}
\end{equation}%
which reproduces equation of motion (\ref{sineG}). The boundary conditions
for the function $H$ are%
\begin{equation}
H(0)=0\ ,\ H(2\pi )=\pm \frac{\pi }{2}\ ,  \label{bc1}
\end{equation}%
which corresponds to $B=\pm 1$ and
\begin{equation}
H(0)-H(2\pi )=0\ ,  \label{bc1.1}
\end{equation}%
which corresponds to $B=0$. The sector $B=0$ is relevant in the construction
of Skyrmion anti-Skyrmion bound states.

Taking into account that the Skyrme model in (3+1) dimensions is the
prototype of non-integrable systems, the above results in Eqs. (\ref{sineG}%
), (\ref{T00a}) and (\ref{LagSG}) are quite remarkable since they show that
the full (3+1)-dimensional Skyrme field equations, energy density and
effective action in a topologically non-trivial sector (as $\rho _{B}\neq 0$%
) can be reduced to the corresponding quantities of the (1+1)-dimensional
sine-Gordon model. The latter is a well known example of integrable models,
see \cite{integrable3} for a detailed review. In particular, it is trivial
to construct kink-like solutions of Eq. (\ref{sineG}) satisfying the
boundary conditions in Eq. (\ref{bc1}) (see \cite{integrable3} and
references therein) and which (due to Eq. (\ref{td1})) represent analytic
(anti)Skyrmions living in the finite flat box defined above.

Since Eqs. (\ref{sineG}), (\ref{T00a}) and (\ref{LagSG}) allow to use all
the available results in Sine-Gordon theory to analyze the (3+1)-dimensional
Skyrme model at finite density, it is useful to follow the conventions of
\cite{coleman}.\ The effective action for a rescaled the Skyrmion profile $%
\Phi $ is%
\begin{equation*}
S=\ell ^{3}\int \!\!d\gamma d\phi \int \!\!dtdr\mathcal{L\ }(\Phi ),\ \quad
H=\frac{\ell }{(\lambda +2\ell ^{2})^{1/2}}\Phi \ ,
\end{equation*}%
where $\ell ^{3}$\ comes from the square root of the determinant of the
metric. Thus the effective Lagrangian for $\Phi $\ reads%
\begin{equation}
\mathcal{L}(\Phi )=-\frac{1}{2}\nabla ^{\mu }\Phi \nabla _{\mu }\Phi +\frac{%
\alpha }{\beta ^{2}}\left( \cos \left( \beta \Phi \right) -1\right) \ ,
\label{SG1}
\end{equation}%
\begin{equation}
\alpha =\frac{\lambda }{2\ell ^{2}\left( \lambda +2\ell ^{2}\right) },\quad
\beta =\frac{4\ell }{\sqrt{\lambda +2\ell ^{2}}}\ .  \label{SG2}
\end{equation}%
The effective sine-Gordon coupling constant that appears from the Skyrme
model always satisfies the Coleman bound $\beta ^{2}<8\pi $.

The mapping presented above allows to construct analytic
Skyrmion--anti-Skyrmion bound states. Namely, the breather-like solutions of
Eq. (\ref{sineG}) satisfying the boundary conditions in Eq. (\ref{bc1.1})
(which correspond to kink anti-kink bound states) correspond\footnote{%
Indeed, under the present mapping from the (3+1)-dimensional Skyrme model
into the (1+1)-dimensional Sine-Gordon model, the (anti)Skyrmion is mapped
into the (anti)kink. Consequently, kink-antikink bound states correspond to
Skyrmion-antiSkyrmion bound states.} to \textit{analytic
Skyrmion--anti-Skyrmion bound states}. To the best of authors knowledge,
this is the first analytic construction of Skyrmions--anti-Skyrmions bound
states in the original (3+1)dimensional Skyrme model. In particular the
number $n_{B}$ of bound states satisfies $n_{B}\leq \frac{8\pi }{\beta ^{2}}%
-1$. Note that already Skyrme and Perring \cite{SkPer} used sine-Gordon in $%
1+1$ dimensions as a \textquotedblleft toy model" for the $3+1$ dimensional
Skyrme model. What is remarkable about the present treatment is that we
found a nontrivial topological sector of the full Skyrme model in which they
are \textit{exactly equivalent}.

The semi-classical quantization in the present sector of the Skyrme model
can be analyzed following \cite{ANW} \cite{bala0}. One first has to identify
the (classical) low energy modes and then it is necessary to quantize such
modes. In the present case, the task is simplified by one of the results
mentioned above: namely, not only the Skyrme field equations with the
generalized hedgehog ansatz in Eqs. (\ref{pions1}), (\ref{pions2}), (\ref%
{pions2.25}) and (\ref{pions2.26}) reduce to the sine-Gordon equation but
also the full Skyrme action reduces to the corresponding sine-Gordon action
in $1+1$ dimensions with the coupling constants defined in Eq. (\ref{SG2}).
Thus, the \textit{principle of symmetric criticality} \cite{palais} applies
in the present case. Consequently, the low energy semi-classical
fluctuations of the Skyrme model in the sector described by Eqs. (\ref%
{pions1}), (\ref{pions2}), (\ref{pions2.25}) and (\ref{pions2.26}) are
described by the reduced action itself (which is nothing but the sine-Gordon
action). Thus, all the known semi-classical results on the sine-Gordon
theory hold.

\subsubsection{An interesting function}

Here we consider an interesting function $\Delta $ of the Skyrmions with
charge $\pm 1$ defined above which encodes the information about how close
they can get to saturate the Skyrme-BPS bound (which, as already emphasized,
cannot be saturated on flat space-times). Nevertheless, it is interesting to
analyze the following function defined\footnote{%
Only in this subsection, we will adopt the convention that $K=2$ and $%
\lambda =1$ which means, roughly (see page 25 of \cite{skyrev1}, taking into
account that the authors use the opposite convention for the space-time
metric with respect to our), that we are measuring lengths in $fm$ and
energy in $MeV$.} as%
\begin{equation}  \label{bound}
\Delta =E- 12 \sqrt{2} \pi^2 \left\vert B\right\vert =E-12 \sqrt{2} \pi^2
\end{equation}%
where $E$ is the energy of the (anti)Skyrmion defined above and $B$ is its
baryon charge. This relation is nothing but the Bogomol'ny bound, as can be
found in \cite{Zahed}, expressed in our conventions. It is worth to
emphasize that in the usual case of the spherical Skyrmion found by Skyrme
himself, the energy exceeds the topological charge by 23\%.

In this case, the above difference $\Delta $ for static configurations $%
H=H(r)$ can be evaluated explicitly in terms of elliptic integral as
follows. From Eq. (\ref{T00a}) one gets the following expression for the
energy-density%
\begin{equation}
T_{00}=\frac{K}{64\,\ell ^{4}}\left[ 16(\lambda +2\,\ell ^{2})(\partial
_{r}H)^{2}+\lambda \left( 1-\cos (4H)\right) +16\,\ell ^{2}\right] \,.
\label{enden1}
\end{equation}%
The field equations Eq. (\ref{sineG}) can be reduced to%
\begin{equation}
\left( 1+2\,\ell ^{2}\right) \frac{\left( H^{\ \prime }\right) ^{2}}{2}+%
\frac{1}{32}\cos (4H)=I_{0}\ \Rightarrow \ \
\end{equation}
\begin{equation}
\frac{dH}{dr}=\pm \sqrt{Q(H)}=\pm \frac{1}{\sqrt{1+2\,\ell ^{2}}}\left(
2I_{0}-\frac{\cos 4H}{16}\right) ^{1/2}\,.  \label{usenerg}
\end{equation}%
In the above equations $I_{0}$ is an integration constant satisfying the
following condition:%
\begin{equation}
\int_{0}^{\pi /2}\frac{dH}{Q(H)^{1/2}}=\pm 2\pi \ ,  \label{usenerg1}
\end{equation}
arising from the requirement to have Baryon charge $\pm 1$ (in this
subsection, from now on we will consider the $+$ sign). The above condition
fixes the integration constant $I_{0}$ as a function of $\ell $:%
\begin{equation}  \label{ellipt}
\int_{0}^{\pi /2}\frac{dH}{Q(H)^{1/2}}=2\pi \ \Rightarrow \
I_{0}=I_{0}\left( \ell \right) .
\end{equation}
An explicit expression however cannot be extracted since,
\begin{equation}  \label{ellipt2}
\int_{0}^{\pi /2}\frac{dH}{Q(H)^{1/2}} = 2\sqrt{2} (1+2 \ell^2) x
K\left(-x^2\right)
\end{equation}
where $x^2 = \frac{2}{32 I_0-1}>0$ and $K(x)$ is the complete elliptic
integral of the first kind.

We can evaluate the energy using Eq. (\ref{usenerg1})
\begin{equation}
E=\pi ^{2}\int_{0}^{\pi /2}\frac{dH}{Q\left( H\right) ^{1/2}}\left( 4\frac{%
(1+2\,\ell ^{2})}{\ell}Q(H)+\frac{\left( 1-\cos (4H)\right) }{4\,\ell}%
+4\ell\right) =E\left( \ell \right) \,.  \label{usenerg2}
\end{equation}

Thus, the natural question is: how far is the energy of the Skyrmions from
saturating the topological bound? The answer depends on the size of the box $%
\ell $. From Eqs. (\ref{usenerg1}) and (\ref{usenerg2}) one can write the
energy $E\left( \ell \right) $ as combinations of elliptic integrals.

\begin{figure}[h]
\begin{subfigure}[t]{0.5\textwidth}
  \centering
  \includegraphics[scale=0.35]{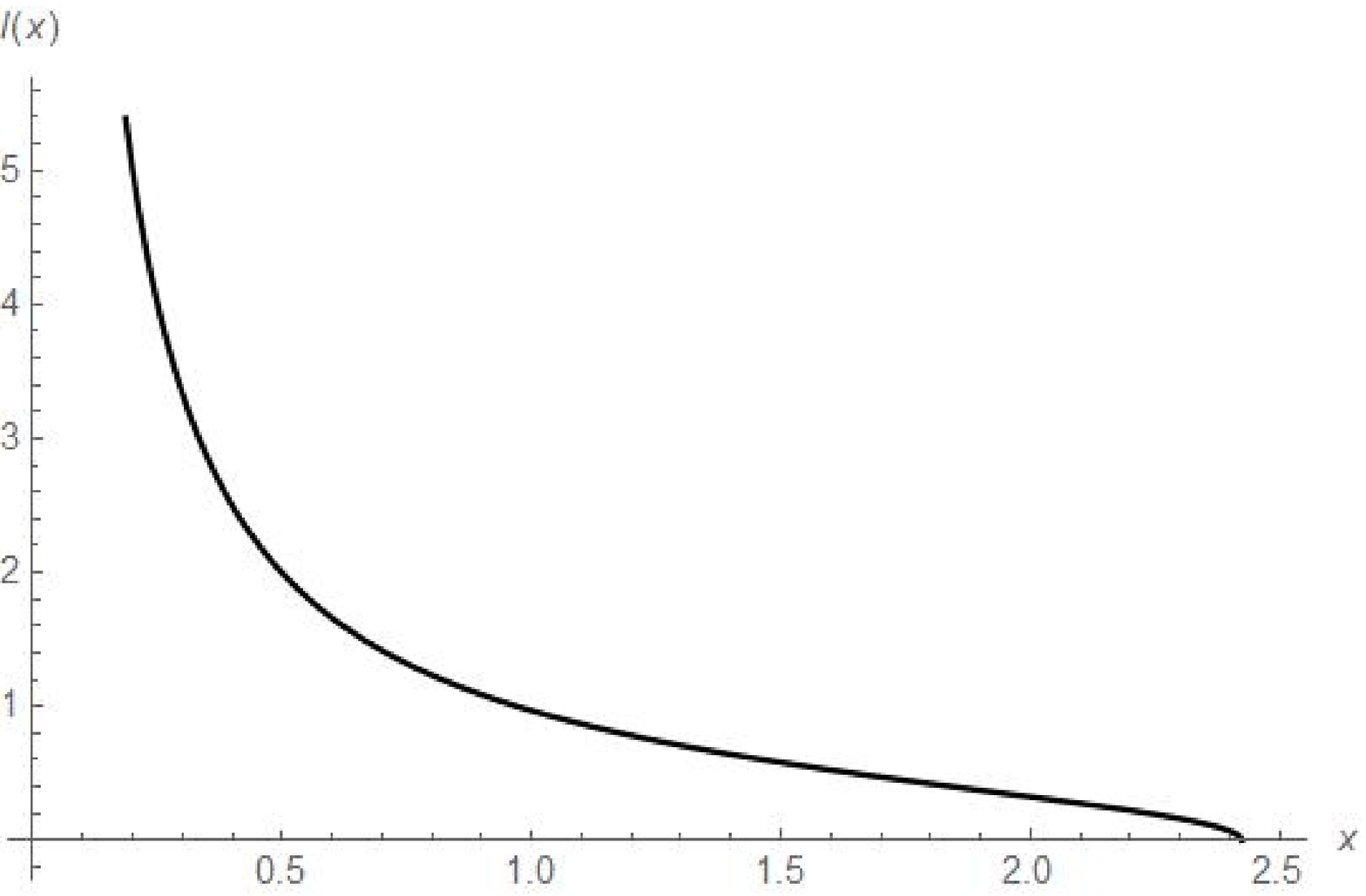}
  \caption{$\ell(x)$}
  \end{subfigure}
\begin{subfigure}[t]{0.5\textwidth}
  \centering
  \includegraphics[scale=0.35]{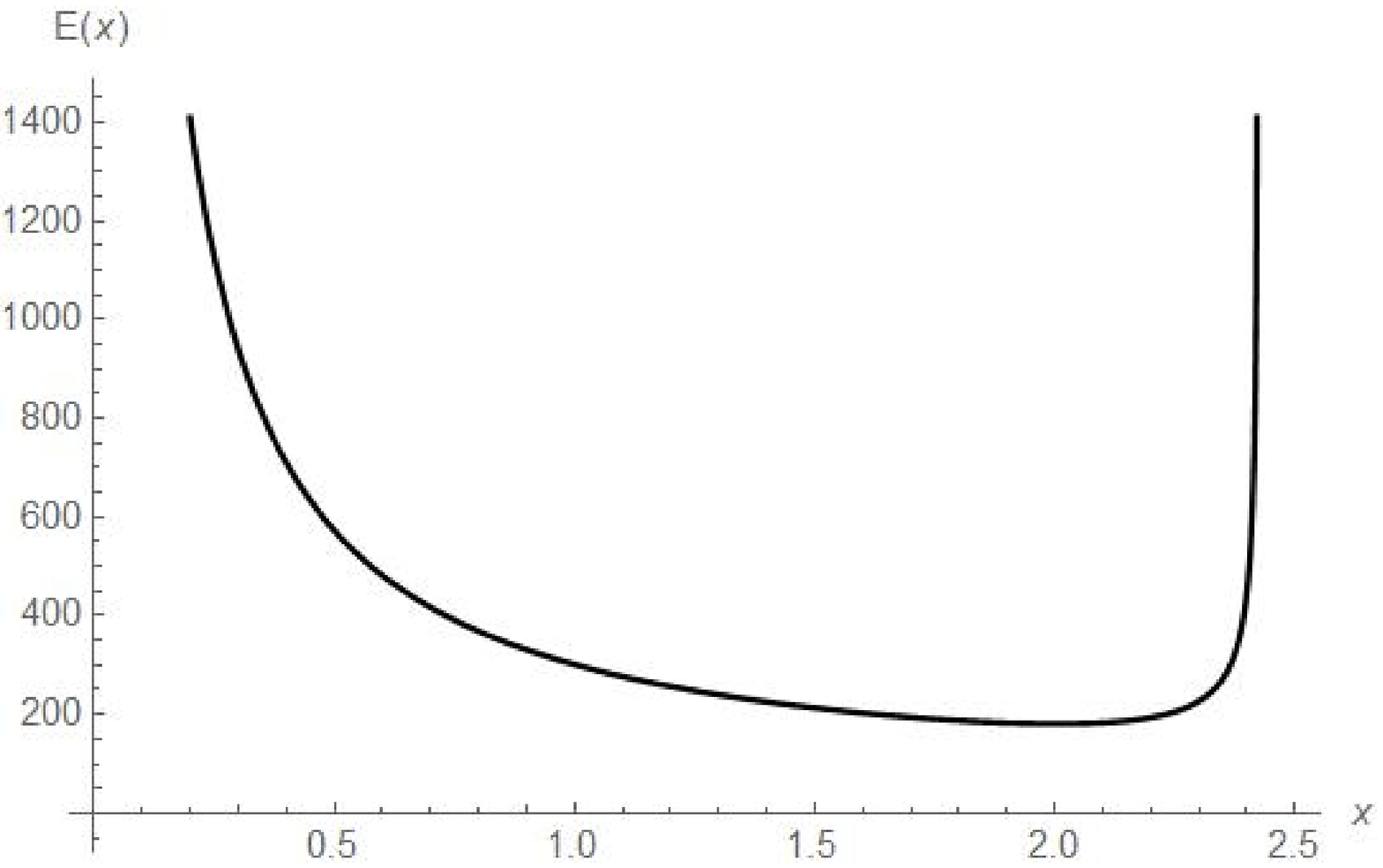}
  \caption{$E(x)$}
  \end{subfigure}
\caption{The size of the box $\ell$ and the energy $E$ as functions of $x=
\left( \frac{2}{32 I_0-1}\right)^{1/2}$.}
\label{Fig1}
\end{figure}

Although relation \eqref{ellipt} cannot provide us with a closed form
relation for $I_0(\ell)$, we can invert it and express $\ell$ as a function
of $I_0$ or of $x$. In this manner we can also express the energy with
respect to $x$. Thus we obtain the two graphs that we see in Fig. \ref{Fig1}%
. From the first graph we can see that $\ell$ and $x$ are in inverse
proportion with each other. At the same time there is a cutoff at a finite
value of $x$, this is owed to taking a finite bound in the integration of
the $r$ variable in \eqref{usenerg1}. As can be seen, by combining %
\eqref{ellipt} and \eqref{ellipt2} to solve algebraically for $\ell$; for $x$
above a certain value, $\ell$ becomes imaginary. The extension of
integration over the full real line should be seen as a pushing of this
bound to infinity. It can be checked arithmetically that the most economic
energy configuration corresponds to a size of the box of $\ell \simeq 0.32$
where $E \simeq 181.312$ which is approximately $8.25\%$ above the lower
bound $12\sqrt{2} \pi^2$ as seen in \eqref{bound}.

These results have a quite natural interpretation. When the size of the box
is small (as the order of $fm$) one should expect strong deviations from
spherical symmetry as extended objects feel strongly the presence of
boundaries. Indeed, the present nonspherical skyrmion is closer to the
topological bound than the usual spherical skyrmion (which exceeds the
topological bound by about $23\%$). On the other hand, when the box size
increases, it should be expected that the presence of the box itself becomes
less relevant. The above plot shows that this is the case since the energy
of the present nonspherical skyrmion grows very rapidly (well above the
topological bound) as the size of the box increases. Thus, when the box is
large enough only small deviations from the spherical skyrmion should be
expected. Consequently, the type of Skyrmions analyzed here is expected to
be favoured at small volumes (or high densities).

\subsection{Inclusion of chemical potential}

The effects of the isospin chemical potential can be taken into account by
using the following covariant derivative (see \cite{chemical1} \cite%
{chemical2})
\begin{equation}
D_{\mu }=\nabla _{\mu }+\bar{\mu}[t_{3},\;\;]\delta _{\mu 0}\ .
\label{newcovdiv}
\end{equation}%
Thus $R_{\mu }$ becomes $\bar{R}_{\mu }=U^{-1}D_{\mu }U$ and the equations
of motion read
\begin{equation}
D^{\mu }\left( \bar{R}_{\mu }+\frac{\lambda }{4}[\bar{R}^{\nu },\bar{G}_{\mu
\nu }]\right) =0\ ,  \label{chemskyrm}
\end{equation}%
where $\bar{G}_{\mu \nu }=[\bar{R}_{\mu },\bar{R}_{\nu }]$. For static
configurations $H(r,z)=H(r)$, the good property of the hedgehog ansatz is
not lost: \textit{the full Skyrme field equations with isospin chemical
potential in Eq. (\ref{chemskyrm}) reduce to just one scalar ODE for} $H$
\begin{equation}
\begin{split}
& \left( \lambda +2\,\ell ^{2}-8\,\lambda \,\ell ^{2}\bar{\mu}^{2}\sin
^{2}(H(r))\right) H^{\prime \prime }(r)-4\lambda \ell ^{2}\bar{\mu}^{2}\sin
(2H(r))H^{\prime 2} \\
& +\lambda \left( \,\bar{\mu}^{2}\ell ^{2}-\frac{1}{8}\right) \sin
(4H(r))+4\,\bar{\mu}^{2}\ell ^{4}\sin (2H(r))=0.
\end{split}
\label{chempotODE}
\end{equation}%
This is a quite non-trivial technical achievement in itself (see, for
instance, \cite{chemical3} \cite{chemical4}). Moreover, the above
differential equation can be reduced to%
\begin{equation}
Y\left( H\right) \frac{\left( H^{\ \prime }\right) ^{2}}{2}+V\left( H\right)
=E_{0}\ ,\ \   \label{chempotODE1}
\end{equation}%
where
\begin{eqnarray*}
Y\left( H\right) &=&\lambda +2\,\ell ^{2}-8\,\lambda \,\ell ^{2}\bar{\mu}%
^{2}\sin ^{2}(H),\  \\
V\left( H\right) &=&-\frac{\lambda }{4}\left( \,\bar{\mu}^{2}\ell ^{2}-\frac{%
1}{8}\right) \cos (4H)-2\,\bar{\mu}^{2}\ell ^{4}\cos (2H)\ .
\end{eqnarray*}%
$E_{0}$ is an integration constant to be determined imposing the physical
boundary condition:%
\begin{equation}
\int_{0}^{\pi /2}\frac{\left[ Y\left( H\right) \right] ^{1/2}}{\left[
E_{0}-V\left( H\right) \right] ^{1/2}}dH=\sqrt{2}2\pi \ .  \label{phybc}
\end{equation}%
Thus, we can compute the critical isospin chemical potential $\overline{\mu }%
_{c}$ as the one for which the above boundary condition cannot be satisfied
anymore as $Y$ can be negative\footnote{%
When $Y(H)$ becomes negative, the numerator in the left hand side of Eq. (%
\ref{phybc}) develops an imaginary part which cannot be compensated by the
denominator.} when $\bar{\mu}\geq \overline{\mu }_{c}$:%
\begin{equation*}
\left( \overline{\mu }_{c}\right) ^{2}=\frac{\lambda +2\,\ell ^{2}}{%
8\,\lambda \,\ell ^{2}}\ .
\end{equation*}

\section{Time crystals}

\label{time crystal}

Obviously, not any time-periodic solution of the Skyrme model is a
time-crystal. For instance, the Skyrmion--anti-Skyrmion bound state
constructed above are time-periodic. However they are not topologically
protected since, if one `pays' the corresponding binding energies, they
decay into the trivial vacuum.

Here we adopt the flat space line element
\begin{equation}
ds^{2}=-d\gamma ^{2}+\ell ^{2}\left( dz^{2}+dr^{2}+d\phi ^{2}\right) \ ,
\label{metric3}
\end{equation}
where $\gamma $ plays the role of time. We have to make the following
modification to ansatz (\ref{pions2.25}), (\ref{pions2.26})
\begin{equation}
A=\frac{\omega \gamma -\phi }{2}\,, \quad G=\frac{\omega \gamma +\phi }{2}\,,
\label{ansnew}
\end{equation}%
where $0\leq \omega \gamma \leq 4\pi $ and the frequency $\omega $ is
necessary to keep $A$ and $G$ dimensionless.

The adoption of line-element \eqref{metric3} means that in this case the
profile $H$ depends on two space-like coordinates. The Skyrme configurations
$U$ defined in Eqs. (\ref{pions1}), (\ref{pions2}), (\ref{pions2.25}), (\ref%
{pions2.26}) and (\ref{ansnew}) are necessarily time-periodic. The full
Skyrme field equations (\ref{nonlinearsigma1}) reduce in this case to
\begin{equation}
\triangle H-\frac{\lambda \omega ^{2}}{4\left( \ell ^{2}(\lambda \omega
^{2}-4)-\lambda \right) }\sin (4H)=0\ ,  \label{thirdeqH}
\end{equation}%
\begin{equation}
\omega ^{2}\neq \omega _{c}^{2}=\frac{\lambda +4\ell ^{2}}{\ell ^{2}\lambda }%
\ ,  \label{crifre}
\end{equation}%
where $\triangle $\ is the two-dimensional Laplacian in $z$ and $r$. Eq. (%
\ref{thirdeqH}) is the Euclidean sine-Gordon equation\footnote{%
It is worth to note that there is a critical value $\omega _{c}$ for the
frequency $\omega $\ of the time crystal (defined in Eq. (\ref{crifre})) at
which Eq. (\ref{thirdeqH}) becomes degenerate. On the other hand, in the
case of the Skyrmions described in the previous section the theory is
defined for all values of the parameters of the model.}. Exact solutions of
Eq. (\ref{thirdeqH})\footnote{%
Previous literature on the analogies between sine-Gordon and Skyme models
can be found in \cite{sin00,sin0,sin1,sin2} and references therein. As it
has been emphasized previously, sine-Gordon theory was believed to be just a
\textquotedblleft toy model" for the 3+1 dimensional Skyrme model. In fact,
we proved that in a nontrivial topological sector they exactly coincide.}
can easily be constructed taking, for instance, $H=H\left( r\right) $.

To construct a time crystal configuration, we firstly need to find stable
kinks satisfying Eq. (\ref{thirdeqH}). As in the previous section, it is
useful to obtain the reduced action $\mathcal{L}$ corresponding to the Eq. (%
\ref{thirdeqH}),
\begin{equation}
\mathcal{L}(H)=\nabla _{\mu }H\nabla ^{\mu }H+\frac{\lambda \omega ^{2}}{%
4\left( \ell ^{2}(\lambda \omega ^{2}-4)-\lambda \right) }\sin ^{2}(2H)\ .
\label{tcsg1}
\end{equation}%
This allows to use well known results on quantization of sine-Gordon theory
also in this sector, for instance, a sine-Gordon kink with $H=H\left(
r\right) $.

Secondly, the configuration has to have a non-trivial winding number. The
topological density is given by $\rho _{B}=3\sin (2H)dH\wedge d\left( \omega
\gamma \right) \wedge d\phi $, and thus the winding number can be evaluated
to
\begin{equation}
W=-\frac{\omega }{8\pi ^{2}}\int_{z=\text{const}}\sin (2H)dHd\gamma d\phi
=\pm 1.
\end{equation}%
This is one of the main results of the paper. We have shown that there are
smooth time-periodic regular configurations of the Skyrme model living at
finite volume possessing a non-trivial winding number along a
three-dimensional time-like surface. Since the winding number is invariant
under any continuous deformation, these configurations can only decay into
other configurations which are also time-periodic (as for static
configurations the above winding number vanishes). Thus, the time
periodicity of these configurations is topologically protected. These are
classical \textit{topologically protected} time-crystals in the sense of
\cite{timec2}. Interestingly enough, the principle of symmetric criticality
\cite{palais}\ can be applied to time-crystal as well since, with the above
time-crystal ansatz not only the Skyrme field equations reduce to the
sine-Gordon system but also the full Skyrme action reduces to the
corresponding sine-Gordon action. Thus, the low energy semi-classical
fluctuations around the time-crystals constructed here are described by the
semi-classical analysis of sine-Gordon theory. Hence, well-known results on
sine-Gordon theory suggest that such time-crystals should also be present at
semi-classical level.

\subsection{The chemical potential}

We can introduce the chemical potential as in the previous section. The full
Skyrme field equations with isospin chemical potential reduce to just one
scalar partial differential equation for the profile $H$,
\begin{equation}
\begin{split}
& \left( \lambda +4\ell ^{2}-\lambda \ell ^{2}\omega ^{2}+8\lambda \ell ^{2}%
\bar{\mu}(\omega -2\bar{\mu})\sin ^{2}(H)\right) \triangle H+4\lambda \ell
^{2}\bar{\mu}(\omega -2\,\bar{\mu})\sin (2H)\nabla _{\mu }H\nabla ^{\mu }H \\
& +4\ell ^{2}\bar{\mu}(2\,\bar{\mu}-\omega )\sin (2H)+\frac{\lambda }{4}%
(\omega -2\,\bar{\mu})^{2}\sin (4H)=0\ .
\end{split}
\label{thirdpoteq}
\end{equation}

In this case the critical chemical potential $\overline{\mu }^{\ast }$
corresponding to time crystal can be found easily as in the previous section
(see the comments below Eqs. (\ref{chempotODE}) and (\ref{chempotODE1})). In
particular, let us consider a kink-like solution of Eq. (\ref{thirdpoteq})
in which the profile only depends on one coordinate $H=H(r)$ and satisfying
the boundary condition in Eq. (\ref{bc1}). Then, Eq. (\ref{thirdpoteq}) reads%
\begin{equation}
Y_{1}\left( H\right) \frac{\left( H^{\ \prime }\right) ^{2}}{2}+V_{1}\left(
H\right) =E_{0}\ ,\ \   \label{chempotODE2}
\end{equation}%
where
\begin{eqnarray*}
Y_{1}\left( H\right) &=&\lambda +\ell ^{2}\left[ 4\,-\lambda \omega
^{2}+8\,\lambda \,\bar{\mu}\left( \omega -2\bar{\mu}\right) \sin ^{2}(H)%
\right] \ ,\  \\
V_{1}\left( H\right) &=&-\frac{\lambda }{16}\left( \omega -2\bar{\mu}\right)
^{2}\cos (4H)-2\ell ^{2}\,\bar{\mu}\left( 2\bar{\mu}-\omega \right) \cos
(2H)\ .
\end{eqnarray*}%
Thus, the critical chemical potential $\overline{\mu }^{\ast }$ can be
determined by requiring%
\begin{equation}
\lambda +\ell ^{2}\left[ 4\,-\lambda \omega ^{2}+8\,\lambda \,\overline{\mu }%
^{\ast }\left( \omega -2\overline{\mu }^{\ast }\right) \right] \leq 0\ ,
\label{critime}
\end{equation}%
since, when this happens, the boundary condition in Eq. (\ref{bc1}) cannot
be satisfied anymore. It is also interesting to note that high values
(compared to $\lambda $) of the time-crystal frequency $\omega ^{2}$
decrease the critical chemical potential: hence, low values of $\omega ^{2}$
are favoured from the thermodynamical point of view.

There is a further special value for the chemical potential (which does not
coincide with the one defined in Eq. (\ref{critime})) for time-crystal
configurations. Indeed, for $\overline{\mu }=\omega /2$, the non-linear
partial differential equation Eq. (\ref{thirdpoteq}) reduces to the linear $%
\triangle H=0$, which, obviously, possesses more symmetries than the generic
one for $\overline{\mu }<\omega /2$. Since this value $\overline{\mu }%
=\omega /2$ of the isospin chemical potential corresponds to a symmetry
enhancement of the field equations, it is natural to wonder whether it is
related to some phase transition of the system. We hope to come back on this
interesting issue in a future publication.

\section{Conclusions}

\label{conclusions}

We constructed the first analytic examples of Skyrmions as well as of
Skyrmions--anti-Skyrmions bound states on flat spaces at finite volume. We
have derived an analytic upper bound for the number of
Skyrmion--anti-Skyrmion bound states in terms of the parameters of the
model. The critical isospin chemical potential can also be computed. With
the same formalism, one can build topologically protected time-crystals:
these are exact configurations of the Skyrme model whose time-dependence is
topologically protected by the non-vanishing winding number. We computed the
corresponding critical isospin chemical potential and determined a possible
experimental signature of these time-crystals.

The present construction answers positively to the question posed in \cite%
{timec2} on the existence of a classical time crystal in systems possessing
non-vanishing topological charges. In fact, as these classical
configurations are topologically protected, the presence of quantum
fluctuations cannot destroy them. Using the results presented above it is
easy to show that these configurations are also present in the
semi-classical quantization of the model. The reason why the powerful no-go
theorems in \cite{timec4} \cite{timec5} do not apply in the present case is
that, in these theorems (both explicitly and implicitly) it is assumed that
the ground state of the theory is static. On the other hand, in theories
with non-Abelian internal

symmetries each non-trivial topological sector has its own ground state.

For instance, in the case of non-Abelian gauge theories admitting BPS
monopoles, the ground state in the sector with unit non-Abelian magnetic
charge is the well-known BPS monopole which cannot be deformed continuosly
to the trivial vacuum. Such a ground state is not invariant under the full
symmetry group of the trivial ground state $|0>$. In particular, it is not
invariant under spatial rotations (unless they are compensated by internal
rotations). This fact, is the origin of the \textquotedblleft spin from
isospin effect\textquotedblright\ discovered in the seventies. In our case,
in the sectors we have named time-crystal, the ground state is time-periodic
and consequently the theorems in \cite{timec4} and \cite{timec5} do not
apply, as we discussed above.

In the present paper, we focused on the $SU(2)$ Skyrme model, but our
results can be generalized to any theory with $SU(N)$ internal symmetry.

\subsection*{Acknowledgements}

This work has been funded by the Fondecyt grants 1160137, 1161150, 3150016
and 3160121. The Centro de Estudios Cient\'{\i}ficos (CECs) is funded by the
Chilean Government through the Centers of Excellence Base Financing Program
of Conicyt.

\end{document}